\documentclass[12pt]{article}

\usepackage{color}

\newcommand{\I}{{\cal I}} 
\newcommand{\hj}{{\hat {\j}}} 
 
\newcommand{\cond}{{\bf \Sigma}} 

\title{Long range correlations and phase transition in non-equilibrium diffusive systems}
\date{\today}
\author{T. Bodineau\footnote{Ecole Normale Sup\'erieure, DMA, 45 rue d'Ulm
75230 Paris cedex 05, France},
B. Derrida\footnote{Laboratoire de Physique Statistique
(CNRS UMR 8550), \'Ecole Normale Sup\'erieure, 24 rue Lhomond, 75231 Paris cedex 05, France}
, V. Lecomte\footnote{D\'epartement de Physique de la Mati\`ere Condens\'ee, Universit\'e de Gen\`eve, 24 quai Ernest Ansermet, 1211 Gen\`eve, Switzerland}
and F. van Wijland\footnote{Laboratoire Mati\`ere et Syst\`emes Complexes 
(CNRS UMR 7057), 10 rue Alice Domon et L\'eonie Duquet, 
Universit\'e Paris Diderot, 75205 Paris cedex 13, France}}

\begin{document}

\maketitle

\begin{abstract}
We obtain explicit expressions for the long range correlations in the ABC model and in diffusive models conditioned to produce an atypical current of particles.
In both cases, the  two-point correlation functions allow to detect the occurrence of a phase transition as they become singular when the system approaches the transition.
\end{abstract}


\section{Introduction}

A generic property of non-equilibrium systems, maintained in a steady state by contact with 
several reservoirs of particles at unequal chemical potentials
(or by contact with several heat baths at unequal temperatures),  
 is the presence of long range correlations 
\cite{Spohn,SC,DKS,OS,DELO,BDGJL-cor}. In diffusive systems, these long
range correlations and their dependence on the system size  \cite{Spohn,DLS5}
can be related to  the non-local nature of the large
deviation function of the density \cite{derrida2007}. Another generic
property of non-equilibrium systems is the possibility of phase
transitions for one dimensional systems with short range 
interactions \cite{Evans,KLMST,PFF,EKKM1,EKKM2,CDE,FF1,BD2005,FF2,ADLW,GMS}, 
in contrast with equilibrium systems where such phase transitions are excluded by the well known Landau argument.

The goal of the present paper is to  calculate these long range correlations,  for  two examples of diffusive systems,  and to show that they become singular at the phase transition, even when the analytic expression of quantities such as the average current or its large deviation function has no singularity as one approaches the phase transition. 

The paper is organized as follows: in section 2, we analyze the ABC model
\cite{EKKM1,EKKM2,CDE,FF1,FF2}, for which we compute first the correlation functions in the case of equal densities of the three species, from the known large deviation function of the density.
We then obtain the correlation functions  in the general case, i.e. for
arbitrary densities,   by two methods:  
a simple truncation procedure based on the expected scaling on the system size of the various correlation functions 
and the macroscopic fluctuation  theory  which was used recently by  Bertini,  De Sole, 
Gabrielli,   Jona-Lasinio,   Landim  \cite{BDGJL1,BDGJL2,BDGJL3,BDGJL4}
in the context of non-equilibrium systems.

In section 3 we consider the weakly asymmetric exclusion process, for which  a phase transition is expected to occur, in the large deviation function of the current. We calculate the pair correlation functions, conditioned on a current deviation, and  show that they become singular at the transition.

\section{The ABC model}

The ABC model \cite{EKKM1,EKKM2,CDE,FF1,FF2}   describes a system of three species $A, B, C$ of particles  on a one dimensional  ring of $L$ lattice sites.  Each
site is occupied by one and only one of these three types of particles.  The dynamics are fully specified by the following exchange rates between neighboring sites:
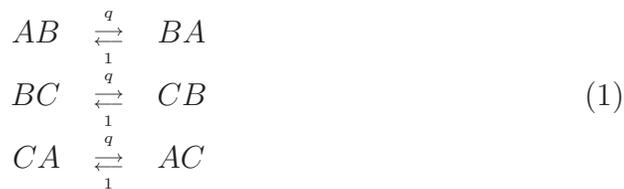
\begin{eqnarray} AB & {{q \atop \longrightarrow}\atop
{\longleftarrow \atop 1}} & BA
\nonumber \\ BC & {{q \atop \longrightarrow}\atop
{\longleftarrow \atop 1}} & CB
\label{update}
\\ CA & {{q \atop \longrightarrow}\atop
{\longleftarrow \atop 1}} & AC
\nonumber
\end{eqnarray}
where $q \leq 1$ (the range $q>1$ is related to the case $q<1$ by a left-right symmetry). 
Since the number of particles of each species is conserved by the dynamics, 
the properties of the system depend on the global densities  $r_A, r_B, r_C$ of the three species and on the parameter $q$. As all sites are occupied by one of the three species, one  has of course
\begin{equation}
r_A+r_B+r_C=1 \  . \label{unity}
\end{equation}
When $q=1$ all allowed configurations are equally likely  (and the steady state density profiles of the three species are flat) but as $q$ decreases, the different species tend to 
segregate  and in the limit $q \to 0$ the steady state consists of three clusters: a cluster of all $A$'s  followed by a cluster of all $B$'s which is itself followed by a cluster of all $C$'s.

Let us first briefly recall some known properties of the ABC model \cite{CDE}.
As $q $ varies, with the following scaling depending on the size $L$ of the ring,
\begin{equation}
q =\exp \left[ - {\beta \over L} \right]  
\label{q-scaling}
\end{equation}
  one observes (for large $L$) phase transitions, in the steady state, at some critical value $\beta_c$.

For equal densities ($r_A=r_B=r_C = {1 \over 3}$), it is known that 
\begin{equation}
\beta_c = 2 \pi \sqrt{3} 
\label{betac-equal}
\end{equation} and  that  for $\beta< \beta_c$ the density profiles of
the three species are flat, while they become space-dependent for $\beta>
\beta_c$.

A stability analysis of the flat phase leads to the following prediction of
$\beta_c$ in the general case
\begin{equation}
\beta_c=
 \frac{ 2\pi}{\left[ 1 - 2 (r_A^2 + r_B^2 + r_C^2)\right]^{1/2}}
\label{betac-unequal}
\end{equation}
which should be the exact expression as long as the transition remains a second order phase transition.
One expects, however, a first order phase transition to occur  
for some range of densities at least when $r_A^2 + r_B^2 + r_C^2  < 2 (r_A^3 + r_B^3 + r_C^3)$,
and the precise location  of this first order transition is not known.
In the flat phase, the  steady state currents  have expressions
$$J_A =  \beta {r_A (r_C - r_B) \over L}$$
(and similar expressions for $J_B$ and $J_C$) 
which do not show any singularity as $\beta \to \beta_c^{-}$.
 
In the steady state, the probability  of observing macroscopic profiles
$\rho_A(x),\rho_B(x), \rho_C(x)$  (where $0 \leq x < 1$ is a macroscopic
coordinate along the ring)  has the following large $L$ dependence
\begin{equation}
 {\rm Pro} (\rho_A(x),\rho_B(x),\rho_C(x)) \sim \exp \Big( - L {\cal F}[\rho_A(x), \rho_B(x), \rho_C(x)] \Big) \, .
\label{F-def}
\end{equation}
The  large deviation function ${\cal F}$ of the density
profiles is  known for the ABC model when $r_A=r_B=r_C = {1 \over 3}$:
\begin{eqnarray}
\label{ld1}
&& {\cal F}[\rho_A(x), \rho_B(x), \rho_C(x)] = \\  
&& \qquad \kappa +  \int_{0}^{1} dx \left[ \rho_A(x) \ln \rho_A(x) +\rho_B(x) \ln \rho_B(x)
+\rho_C(x) \ln \rho_C(x) \right] \hspace{0.5cm} \nonumber\\
&& \qquad +    \beta \int_{0}^{1} dx \int_{0}^{1}  dz \ z \left[ \rho_B(x) \rho_C(x+z) +
\rho_C(x) \rho_A(x+z) + \rho_A(x) \rho_B(x+z)\right] \, . \nonumber 
\end{eqnarray}
where $\kappa$ is a normalisation constant but it
is not known for arbitrary $r_A,r_B,r_C$. 
So far it has only  been computed in \cite{CDE} to order $\beta^2$
\begin{eqnarray}
\lefteqn{{\cal F}[\rho_A(x), \rho_B(x), \rho_C(x)] =}
\nonumber \\
&& \kappa 
+  \int_{0}^{1} dx \left[
\rho_A(x) \ln \rho_A(x) +\rho_B(x) \ln \rho_B(x)
+\rho_C(x) \ln \rho_C(x) \right]  \nonumber\\
&& +   \beta \int_{0}^{1} dx \int_{0}^{1}  dz \ z
\left[ \rho_B(x) \rho_C(x+z) +
\rho_C(x) \rho_A(x+z) + \rho_A(x) \rho_B(x+z)\right]  \nonumber \\
&& - {3 \over 4}  \beta^2 \int_{0}^{1} dx \int_0^1 dz \;  z (1-z) \left[
 r_A (1 - 3 r_A) \rho_B(x) \rho_C(x+z)  \right. \nonumber \\
&& \qquad \left.
+  r_B (1 - 3 r_B)\rho_C(x) \rho_A(x+z)
+ r_C(1-3 r_C)\rho_A(x) \rho_B(x+z) \right] \nonumber\\
&& \qquad \qquad \qquad 
+ O( \beta^3)
\label{ld2}
\end{eqnarray}

\subsection{The correlation function in the equal density case}

When the large deviation function is known, as in (\ref{ld1}) for the
equal density case, the calculation of the correlation functions in the
flat phase can be easily done from the contribution of the fluctuations of density around the flat profiles.
For a small perturbation of the form
$$\rho_A(x)= {1 \over 3} + a_1 \cos(2 \pi n x) + a_2 \sin(2 \pi n x)$$
$$\rho_B(x)= {1 \over 3} + b_1 \cos(2 \pi n x) + b_2 \sin(2 \pi n x)$$
with $n  \ge 1$ 
and $\rho_C(x)=1-\rho_A(x)-\rho_B(x)$,
one gets from (\ref{ld1}) at quadratic order in $a_1,a_2,b_1,b_2$
\begin{eqnarray*}
&& {\rm Pro} (\rho_A(x),\rho_B(x),\rho_C(x)) \sim \\
&&  \qquad
\exp\left[ -   {3L \over 2} (a_1^2 + a_1 b_1 + b_1^2 + a_2^2 + a_2 b_2 + b_2^2) + {3L  \beta \over 4 n \pi} (a_1 b_2 - a_2 b_1) \right] 
\end{eqnarray*}
Therefore,  the covariances  in the steady state of the $n$-th Fourier coefficients of the
densities $\rho_A(x)$ and $\rho_B(x)$ are
$$ \langle a_1^2 \rangle =
 \langle a_2^2 \rangle =
 \langle b_1^2 \rangle =
 \langle b_2^2 \rangle = {16 n^2 \pi^2 \over 3 L (12 n^2 \pi^2 - \beta^2)} $$
$$ \langle a_1 a_2 \rangle =  \langle b_1 b_2 \rangle = 0$$
 $$\langle a_1b_1 \rangle =
 \langle a_2b_2 \rangle = -{8 n^2 \pi^2 \over 3 L (12 n^2 \pi^2 - \beta^2)} $$
$$\langle a_1 b_2 \rangle =  - \langle a_2 b_1 \rangle ={ 4 \pi n \beta \over 3 L (12 n^2 \pi^2 - \beta^2)}$$
Then if one sums over all the fluctuations, i.e. over all the wave numbers $n$ on the ring, one obtains
\begin{eqnarray}
\langle\rho_A(x) \rho_A(y) \rangle_c= {16  \pi^2 \over 3 L } \sum_{n \geq 1} {n^2 \cos(2\pi n (y-x)) \over (12 n^2 \pi^2 - \beta^2)} 
\label{rara1}
\end{eqnarray}
\begin{eqnarray}
\langle\rho_A(x) \rho_B(y) \rangle_c= {-8  \pi^2 \over 3 L } \sum_{n \geq 1} {n^2 \cos(2\pi n (y-x)) \over (12 n^2 \pi^2 - \beta^2)}+
  {4 \pi \beta \over 3 L}  \sum_{n \geq 1} {n \sin(2\pi n (y-x)) \over (12 n^2 \pi^2 - \beta^2)}
\label{rarb1}
\end{eqnarray}
and similar expressions for the other correlation functions
(we will see below in (\ref{rara4},\ref{rarb4})  a different way of writing these expressions).
Clearly these expressions become singular as $\beta$ approaches the
critical value  $\beta_c=2 \pi \sqrt{3}$.

\subsection{The correlation function for arbitrary densities by a simple truncation procedure}
\label{sec: truncation}

Let us introduce the notation $A_i=1$ when site $i$ is occupied by a $A$
particle, $A_i=0$ otherwise, and similarly $B_i$ or $C_i$ for a $B$ or a $C$ particle,  so that on each site $i$ one has
\begin{equation}
A_i+ B_i + C_i = 1 \, .
\label{sum-rule}
\end{equation}
One can  derive from the dynamical rule (\ref{update}) the following exact evolution equations:
\begin{eqnarray}
{d  \langle A_i \rangle \over dt}=
q \langle A_{i-1} B_i  \rangle + \langle B_i A_{i+1}   \rangle
-q \langle A_{i} B_{i+1}  \rangle - \langle  B_{i-1}  A_i \rangle
\nonumber
\\
+  \langle A_{i-1} C_i  \rangle + q \langle C_i A_{i+1}  \rangle
- \langle A_{i} C_{i+1}  \rangle - q \langle C_{i-1} A_{i}   \rangle \, ,
\label{evo1}
\end{eqnarray}
where  $\langle \ \rangle$ stands for the expectation with respect to the dynamics.
For $|j-i| \ge 2$
\begin{eqnarray}
{d  \langle A_i  A_j\rangle \over dt}=
 q \langle A_{i-1} B_i A_j  \rangle + \langle  B_i A_{i+1}  A_j \rangle
-q \langle A_{i} B_{i+1} A_j \rangle - \langle B_{i-1} A_{i}   A_j \rangle
\nonumber
\\
+ q \langle A_i A_{j-1} B_j   \rangle + \langle  A_i B_j A_{j+1}   \rangle
-q \langle A_{i} A_j B_{j+1}  \rangle - \langle  A_{i}  B_{j-1} A_j \rangle
\nonumber
\\
+  \langle A_{i-1} C_i A_j  \rangle +  q \langle  C_i A_{i+1}  A_j \rangle
- \langle A_{i} C_{i+1} A_j \rangle -  q \langle C_{i-1} A_{i}   A_j
\rangle
\nonumber \\
+  \langle A_i A_{j-1} C_j   \rangle +  q \langle  A_i C_j A_{j+1}   \rangle
- \langle A_{i}  A_j C_{j+1}  \rangle -  q \langle  A_i C_{j-1}    A_j \rangle
 \label{evo2}
\end{eqnarray}
\begin{eqnarray}
{d  \langle A_i  B_j\rangle \over dt}=
 q \langle A_{i-1} B_i B_j  \rangle + \langle  B_i A_{i+1}  B_j \rangle
-q \langle A_{i} B_{i+1} B_j \rangle - \langle B_{i-1} A_{i}   B_j \rangle
\nonumber
\\
+ q \langle A_i B_{j-1} C_j   \rangle + \langle  A_i C_j B_{j+1}   \rangle
-q \langle A_{i} B_j C_{j+1}  \rangle - \langle  A_{i}  C_{j-1} B_j \rangle
\nonumber
\\
+ \langle A_{i-1} C_i B_j \rangle + q \langle  C_i A_{i+1} B_j \rangle
- \langle A_i C_{i+1} B_j \rangle - q \langle  C_{i-1} A_i B_j \rangle
\nonumber
\\
+  \langle A_i B_{j-1} A_j   \rangle +  q \langle  A_i A_j B_{j+1}   \rangle
- \langle A_{i} B_j A_{j+1}  \rangle -  q \langle  A_i A_{j-1}   B_j
\rangle
\label{evo3}
\end{eqnarray}
and for $j=i+1$:
\begin{eqnarray}
{d  \langle A_i  B_{i+1}\rangle \over dt}=
 q \langle A_{i-1} B_i B_{i+1}  \rangle + \langle  B_i A_{i+1}  \rangle
-q \langle A_{i} B_{i+1}  \rangle - \langle B_{i-1} A_{i}   B_{i+1}
\rangle
 \nonumber \\
 + \langle  A_i C_{i+1} B_{i+2}
\rangle
-q \langle A_{i} B_{i+1} C_{i+2}  \rangle
 \nonumber \\
+ \langle A_{i-1} C_i B_{i+1} \rangle
- q \langle  C_{i-1} A_i B_{i+1} \rangle
 +  q \langle  A_i A_{i+1} B_{i+2}
\rangle
- \langle A_{i} B_{i+1} A_{i+2}  \rangle
\nonumber \\
\label{evo5}
\end{eqnarray}
and similar expressions for the evolution of the other one or two point functions. 
In the steady state, the left hand sides of the equations (\ref{evo1}-\ref{evo5}) vanish.

For  diffusive systems, one expects the long range part of the $p$-point
connected part of the correlation functions to scale as  $L^{1-p}$ in the
steady state \cite{Spohn,DLS5,derrida2007}.
This means that, for large $L$,  in the flat phase of the $ABC$ model,
\begin{equation}
\langle A_i B_{j}   \rangle  = \langle A_i    \rangle \langle 
B_{j}   \rangle +\langle A_i B_{j}   \rangle_c = r_A r_B + {1 \over L} F_{ab}\left( {j-i
\over L} \right) 
\label{sca2}
\end{equation}
where $F_{ab}$ stands for the normalized connected two-point function and
\begin{eqnarray}
\langle A_i B_j C_k   \rangle  = r_A r_B r_C &+&
 {1 \over L} \left[  r_C F_{ab}\left( {j-i \over L} \right)
+ r_B F_{ac}\left( {k-i \over L}\right)
+ r_A F_{bc}\left( {k-j \over L} \right)
\right]
\nonumber
\\ 
\label{scaling}
 &+&{1 \over L^2}  G_{abc}\left( {j-i \over L}, {k-i \over L}\right)
\label{sca3}
\end{eqnarray}
where $G_{abc}$ stands for the normalized connected three-point function.
Using similar notations for all the other 2-point and 3-point
functions and the relation (\ref{sum-rule}), and replacing the various correlation functions in (\ref{evo2},\ref{evo3},\ref{evo5})  by their scaling forms (\ref{sca2},\ref{sca3}), one gets to leading orders in $L^{-1}$ (order $L^{-3}$ for (\ref{evo2},\ref{evo3}),
and order $L^{-1}$ for  (\ref{evo5}))
\begin{eqnarray}
&&  F_{aa}(x)''= \beta r_A [F_{ba}(x)'- F_{ab}(x)'] \nonumber \\
&& F_{bb}(x)''= \beta r_B [F_{ba}(x)'- F_{ab}(x)']  \label{equa-dif} \\
&& F_{ab}(x)''= \beta  [r_B F_{aa}(x)' + r_A F_{bb}(x)' - (1-2 r_A - 2 r_B) F_{ab}(x)'] \nonumber \\
&&  F_{ba}(x)''= \beta  [-r_B F_{aa}(x)' - r_A F_{bb}(x)' + (1-2 r_A - 2 r_B) F_{ba}(x)'] \nonumber 
\end{eqnarray}
as well as
\begin{equation}
F_{ab}(0) -F_{ba}(0) = 3 \beta r_A r_B r_C \, .
\label{eq-bord}
\end{equation}
Note that in (\ref{equa-dif}) at leading  order in $1/L$, the contribution of the functions $G$'s which appear in (\ref{sca3}) drops out.
Note also that $$F_{ab}(x)=F_{ba}(1-x) \ .$$

The general solution of  (\ref{equa-dif}) is 
\begin{eqnarray}
&& F_{aa}(x)= K_{aa} - 2 K \beta r_A \cos \left( {\beta \sqrt{\Delta} \over 2} ( 2 x -1) \right)
\nonumber \\ 
&& F_{bb}(x)= K_{bb} - 2 K \beta r_B \cos \left( {\beta \sqrt{\Delta} \over 2} ( 2 x -1) \right)
\label{F-results} \\
&& F_{ab}(x)= F_{ba}(1-x) =K_{ab}
 + K (1- 2 r_C) \beta  \cos \left( {\beta \sqrt{\Delta} \over 2} ( 2 x -1) \right)
\nonumber \\
&& \qquad \qquad \qquad \qquad  -  K \beta   \sqrt{\Delta}  \sin \left( {\beta \sqrt{\Delta} \over 2} ( 2 x -1) \right)
\nonumber
\end{eqnarray}
where $K, K_{aa},K_{ab}, K_{bb}$ are integration constants  and
\begin{equation}
\Delta = 1 - 2 (r_A^2 + r_B^2 + r_C^2) \, .
\label{Delta-def}
\end{equation}
Relation (\ref{eq-bord}) leads to 
\begin{equation}
K= - {3 r_A r_B r_C \over 2   \sqrt{  \Delta} \sin{ \beta \sqrt{\Delta} \over 2}} \, .
\end{equation}
The remaining constants can be determined by using the fact that
$$ \sum_{k=1}^{L-1} \langle A_i A_{i+k} \rangle = { N_A (N_A-1) \over L}$$
and similar sum rules for the other correlation functions
which imply   that
$$\int_0^1 F_{aa}(x) dx = - r_A (1- r_A), \ 
\int_0^1 F_{bb}(x) dx = - r_B (1- r_B), \ 
\int_0^1 F_{ab}(x) dx =  r_A r_B $$
 and the final result is for $0 < x < 1$
\begin{equation}
F_{aa}(x) = - r_A (1- r_A) - {3 r_A^2 r_B r_C \over   \sqrt{\Delta} \sin{ \beta \sqrt{\Delta} \over 2}}
  \left(
   \beta  \cos \left( {\beta \sqrt{\Delta} \over 2} ( 2 x -1) \right)
- 2   {\sin\left( {\beta \sqrt{\Delta} \over 2 }\right) \over \sqrt{\Delta} }\right)
\label{rara3}
\end{equation}
\begin{eqnarray}
F_{ab}(x) =  r_A r_B +
 {3 r_A r_B r_C \over 2  \sqrt{\Delta} \sin{ \beta \sqrt{\Delta} \over 2}}
  \left[
    (1 - 2 r_C) \left( \beta  \cos \left( {\beta \sqrt{\Delta} \over 2} ( 2 x -1) \right)
- 2   {\sin\left( {\beta \sqrt{\Delta} \over 2 }\right) \over
  \sqrt{\Delta} } \right) 
\right. \nonumber \\
 \left. -  \beta \sqrt{\Delta} \sin \left( {\beta \sqrt{\Delta} \over 2} ( 2 x -1)
   \right)\right]
\label{rarb3}
\end{eqnarray}

As $\beta \to {2 \pi \over \sqrt{\Delta}}$ the correlation functions diverge and one recovers the expression (\ref{betac-unequal}) of the second order transition. 

As starting from (\ref{sca2},\ref{sca3})  we only considered the correlations in  the flat phase, our truncation procedure is unable 
to distinguish between a stable and a metastable phase and to   predict the occurrence of first order transitions. Thus  
 we cannot exclude that the flat phase becomes metastable for some values of $\beta$ smaller than $\beta_c$ given by  (\ref{betac-unequal}).

\subsection{The macroscopic fluctuation theory }

In the $ABC$ model there is always one and only one particle per
site. One can then describe the system on a macroscopic scale and on a diffusive
time, i.e. on times which scale as $L^2$, by only two  density profiles
$\rho_A(x,\tau), \rho_B(x,\tau)$.

The key quantities to the macroscopic description of the system are the typical currents for given density profiles and their variance in the steady state. Let $Q^A_t (i), Q^B_t (i)$ be the fluxes of $A,B$ particles between sites $i$ and $i+1$
during a long microscopic time $t$. Starting at time 0 from smooth macroscopic profiles $\rho_A (x), \rho_B (x)$, the densities  evolve according to \cite{CDE}
\begin{eqnarray}
\label{eq: hydro}
\partial_\tau \rho_A(x,\tau) =  \partial_x^2 \rho_A (x,\tau)  + \beta \partial_x \big( \rho_A (x,\tau)(\rho_B (x,\tau)- \rho_C(x,\tau)) \big)\\
\partial_\tau \rho_B(x,\tau) =  \partial_x^2 \rho_B (x,\tau)  + \beta \partial_x \big( \rho_B (x,\tau)(\rho_C (x,\tau)- \rho_A(x,\tau)) \big)
\nonumber
\end{eqnarray}
where the macroscopic time $\tau$ scales like $t/L^2$.
One can understand (\ref{eq: hydro}) from (\ref{evo1}) by saying that for large $L$, there is a local equilibrium at position  $x$, characterized by
the densities $\rho_A(x,\tau)$ and $\rho_B(x,\tau)$. In this local equilibrium, the hydrodynamic currents are given by
\begin{eqnarray}
 q_A= -{d \rho_A \over dx} - \beta \rho_A(\rho_B- \rho_C) \label{qaqb}\\ 
 q_B= -{d \rho_B \over dx} - \beta \rho_B(\rho_C- \rho_A)
\nonumber
\end{eqnarray} 
To compute the covariance matrix of the currents, it is enough to consider the system in equilibrium with $\beta = 0$ and constant densities $\rho_A,\rho_B$. In this case, the $A$-particles evolve as a Symmetric Simple Exclusion Process (SSEP) (the $B,C$ particles play the role of holes) and the variance of the total current $Q^A_t = \sum_{i=1}^L Q^A_t (i)$  is given by \cite{BD2004}
\begin{eqnarray*}
\sigma_{AA} = \lim_{t \to \infty} \frac{1}{t L} \langle (Q^A_t)^2 \rangle = 2  \rho_A ( 1 - \rho_A) \, .
\end{eqnarray*} 
In the same way, the variance of $Q^B_t = \sum_{i=1}^L Q^B_t (i)$  is given by $\sigma_{BB} = 2  \rho_B ( 1 - \rho_B)$.
Finally, if we view the $C$-particles as holes and the species $A,B$ as a single type of particle, then the evolution is the same as the  SSEP at density $\rho_A + \rho_B$, so that 
\begin{eqnarray*}
\lim_{t \to \infty} \frac{1}{t L} \langle (Q^A_t + Q^B_t)^2 \rangle = 2  (\rho_A + \rho_B) ( 1 - \rho_A - \rho_B) 
= \sigma_{AA} + \sigma_{BA} + 2 \sigma_{AB}  \, .
\end{eqnarray*} 
where the covariance between the currents of $A$ and $B$ is then $\sigma_{AB} = - 2 \rho_A \rho_B$.

\medskip

One can adapt the macroscopic fluctuation theory \cite{HS,KOV}
 to predict that the probability of observing density profiles
$\rho_A(x,\tau), \rho_B(x,\tau)$
and (rescaled) currents $j_A(x,\tau),j_B(x,\tau)$ over a (rescaled) time interval 
$0 < \tau < T$ is given (this rescaling is discussed for example in section 5 of  \cite{BD2007}) 
\begin{equation}
 {\rm Pro}(\rho_A,\rho_B,j_A,j_B) \sim \exp \left( - L  \ {\cal I}_{[0,T]} (\rho_A,\rho_B,j_A,j_B)  \right) \, ,
\label{pro}
\end{equation}
where 
\begin{equation}
\label{eq: I ABC}
{\cal I}_{[0,T]} (\rho_A,\rho_B,j_A,j_B) = \int_0^1 dx
\int_0^T d \tau   \ {\cal H} (\rho_A,\rho_B,j_A,j_B) \, ,
\end{equation}
with 
\begin{eqnarray}
\label{H-def}
&& {\cal H} (\rho_A,\rho_B,j_A,j_B) = \frac{1}{2}
\left( \begin{array}{c} 
j_A - q_A    \\
j_B - q_B  
\end{array}  \right)
\left( \begin{array}{cc} 
\sigma_{AA}      &  \sigma_{AB} \\
 \sigma_{AB}  & \sigma_{BB}
\end{array}  \right)^{-1}
\left( \begin{array}{c} 
j_A - q_A    \\
j_B - q_B  
\end{array}  \right) \\
&& \qquad \qquad = {\sigma_{BB} (j_A - q_A)^2 - 2 \sigma_{AB}(j_A- q_A)(j_B-q_B) +
\sigma_{AA} (j_B - q_B)^2 \over 2 ( \sigma_{AA} \sigma_{BB} - \sigma_{AB}^2)} \nonumber
\end{eqnarray}
Note that instead of (\ref{H-def}), one could  write Langevin equations for the currents
$$j_A= -{d \rho_A \over dx} - \beta \rho_A(\rho_B- \rho_C) + {
 \eta_A(x,\tau) \over \sqrt{L}}$$
$$j_B= -{d \rho_B \over dx} - \beta \rho_B(\rho_C- \rho_A) + {
 \eta_B(x,\tau) \over \sqrt{L}}$$
with the following correlations of the white noises $\eta_A(x,\tau)$ and
$\eta_B(x,\tau)$
$$ \langle \eta_A(x,\tau) \eta_B(x,\tau') \rangle =
\sigma_{AB}(\rho_A(x,\tau),\rho_B(x,\tau)) \  \delta(\tau-\tau')$$
and similar expressions for $\langle \eta_A(x,\tau) \eta_A(x,\tau')
\rangle$ and $\langle \eta_B(x,\tau) \eta_B(x,\tau') \rangle$.

As always, the currents and the densities are related by the conservation laws
$${d \rho_A \over d\tau} = - {d j_A \over dx} 
 \ \ \ \ \ \ ; \ \ \ \ \ {d \rho_B \over d\tau} = - {d j_B \over dx}  \  .$$
For a perturbation of the form (which satisfies these conservation laws)
\begin{eqnarray}
\rho_A(x,\tau)= r_A + k (a_1 \cos[k x + \omega \tau]
+a_2 \sin[k x + \omega \tau]) \nonumber \\
\rho_B(x,\tau)= r_B + k (b_1 \cos[k x + \omega \tau]
+b_2 \sin[k x + \omega \tau]) \nonumber \\
j_A(x,\tau)= - \beta r_A (r_B - r_C) - \omega (a_1 \cos[k x + \omega \tau]
+a_2 \sin[k x + \omega \tau])\nonumber \\
j_B(x,\tau)= - \beta r_B (r_C - r_A) - \omega (b_1 \cos[k x + \omega \tau]
+b_2 \sin[k x + \omega \tau]) 
\label{perturb}
\end{eqnarray}
 where \begin{equation}
k = 2 n \pi \ \ \ \ \ \omega = {2 m \pi \over T}
\label{k-omega}
\end{equation}
with $n \in N$ and $ m \in Z$
 one expands  (\ref{eq: I ABC}-\ref{H-def}) to second order in  $a_1,a_2,b_1,b_2$  and the probability (\ref{pro}) becomes a Gaussian  which  leads to the following expectations 
$$ \langle a_1 a_2\rangle =  \langle b_1 b_2\rangle =  0$$
\begin{eqnarray}
&& \langle a_1^2\rangle = \langle a_2^2\rangle = 
-{4 r_A \over L T }{1 \over  \Gamma_k} 
\Big(  \beta^2 k^2 [4   (1- 4 r_A) r_B r_C-(1-2 r_A)^2 (1-r_A) ]
\nonumber \\
&& \qquad \qquad \qquad  \qquad \qquad   + 2 \beta k (1-2 r_A)(r_C-r_B) \omega -(1-r_A) ( k^4  + \omega^2) \Big)
\nonumber \\
&& \langle a_1 b_1\rangle =  
\langle a_2 b_2\rangle =  
-{4 r_A r_B \over L T}{1 \over  \Gamma_k}  
\Big(  - \beta^2 k^2   [4 r_A r_B  -1 - 2 r_C + 8 r_C^2 ]
\label{eq: quadratic} \\
&& \qquad \qquad \qquad  \qquad \qquad   + 2 \beta k  (r_B-r_A) \omega +  ( k^4  + \omega^2) \Big)
\nonumber \\
&& \langle a_1 b_2\rangle =  
-\langle a_2 b_1\rangle =  
{1 \over L T} \; {24 \beta k^3  r_A r_B  r_C \over \Gamma_k } \nonumber
\end{eqnarray}
where  $\Delta$ is defined in (\ref{Delta-def}) and 
$$
\Gamma_k =   \beta^4 k^4   \Delta^2 - 2 \beta^2 k^2  \Delta ( k^4   - \omega^2) + (k^4  + \omega^2)^2 \, .
$$

\medskip

In one adds up the contributions of all perturbations of the form (\ref{perturb}), one gets for the correlation functions 
$$\langle\rho_A(x) \rho_A(y) \rangle_c=  \sum_{k,\omega}  k^2 \langle a_1^2\rangle    \cos[ k (y-x)]  $$
$$\langle\rho_A(x) \rho_B(y) \rangle_c=  \sum_{k,\omega}  
k^2 \langle b_1^2\rangle    \cos[ k (y-x)] 
+ k^2 \langle a_1 b_2 \rangle    \sin[ k (y-x)] 
 $$

In the long time limit the sums over the discrete frequencies (\ref{k-omega}) can be replaced by an integral and one gets
\begin{eqnarray}
\label{rara2}
&& \langle\rho_A(x) \rho_A(y) \rangle_c= \\
&&  {1 \over  L } \sum_{n \geq 1} 
{ r_A[  \beta^2 ( 3 r_A (1-2 r_A)^2 - \Delta (2- 5 r_A)) +8 \pi^2 n^2 (1-r_A)]
\over   4 \pi^2 n^2  - \beta^2  \Delta}
  \cos(2\pi n (y-x))  
\nonumber 
\end{eqnarray}
\begin{eqnarray}
\langle\rho_A(x) \rho_B(y) \rangle_c= {1 \over  L } \sum_{n \geq 1} 
{ r_A r_B [  2 \beta^2 (\Delta - 3 r_C + 6 r_C^2)) - 8 \pi^2 n^2 ]
\over   4 \pi^2 n^2  - \beta^2  \Delta}
  \cos(2\pi n (y-x))
  \nonumber \\ +
{1 \over  L}  \sum_{n \geq 1} { 12  \beta  \pi n   \ r_A r_B r_C
\over   4 \pi^2 n^2  - \beta^2  \Delta}
\sin(2\pi n (y-x)) 
\nonumber \\ 
\label{rarb2}
\end{eqnarray}
In the particular case of equal densities, one recovers  (\ref{rara1},\ref{rarb1}).
\\ \ \\ \ \\
Using the identities 
\begin{eqnarray}
\label{eq: cos identite}
2 \sum_{n \ge 1} \cos[2 \pi n (x-y)] = -1 + \delta(x-y)
\end{eqnarray}
\begin{eqnarray*}
{\cos[\alpha ( {1 \over 2 }-x)] \over \alpha  \sin{\alpha \over 2}}
 &=& {\cos(\alpha x) + \cos(\alpha(1-x))  \over  \alpha \sin \alpha} = {2 \over
\alpha^2}  -\sum_{n \ge 1} {4 \over 4 \pi^2 n^2 - \alpha^2} \cos{2 n \pi x} \\
{\sin[\alpha ( {1 \over 2 }-x)] \over   \sin{\alpha \over 2}}
&=& {\cos (\alpha x) - \cos(\alpha(1-x)) \over 1- \cos \alpha}
=  \sum_{n \ge 1} {8 n \pi \over 4 \pi^2 n^2 - \alpha^2} \sin{2 n \pi x}
\end{eqnarray*}
$$(1- 2 r_A)^2 + \Delta= 4 r_B r_C$$
one can rewrite (\ref{rara2},\ref{rarb2}) as 
\begin{eqnarray}
\nonumber
 && \langle\rho_A(x) \rho_A(y) \rangle_c=
\\  && {1 \over  L } \left[ r_A (1-r_A) (\delta(y-x)
- 1)  - 3 \beta r_A^2 r_B r_C \left( {\cos\left({ \beta \sqrt{\Delta} \over 2}(1-2 y+ 2
  x) \right)
    \over \sqrt{\Delta} \sin { \beta \sqrt{\Delta} \over 2}} - {2 \over
\beta \Delta} \right)
\right]
\nonumber
\\ \label{rara4}
\end{eqnarray}
\begin{eqnarray}
\nonumber      
 && \langle\rho_A(x) \rho_B(y) \rangle_c=
\\  && {1 \over  L } \left[- r_A r_B (\delta(y-x)
- 1)   +
{3 \over 2} \beta r_A r_B r_C (1- 2 r_C) \left(
 {\cos\left( \beta {\sqrt{\Delta}
\over 2}(1-2 y+ 2
  x) \right)
    \over \sqrt{\Delta} \sin { \beta \sqrt{\Delta} \over 2}} - {2 \over
\beta \Delta} \right)
\right. \nonumber
\\ && \left.
+{3 \over 2} \beta r_A r_B r_C 
 {\sin\left({ \beta \sqrt{\Delta}
\over 2}(1-2 y+ 2 x) \right)  \over  \sin { \beta \sqrt{\Delta} \over 2}} \right]
\label{rarb4}
\end{eqnarray}
which are totally equivalent to the expressions (\ref{rara3},\ref{rarb3}). 

As mentioned before, the large deviation function (\ref{F-def}) of the density is only  known (\ref{ld2})
at order $\beta^2$ for the $ABC$ model. 
The knowledge (\ref{rara4},\ref{rarb4}) imposes constraints on the large deviation function.  Trying to generalize (\ref{ld2}), we found  the following expression for the large deviation functional (\ref{F-def})
\begin{eqnarray}
\label{ld3}
&& {\cal F}[\rho_A(x), \rho_B(x), \rho_C(x)] = \\
&& \kappa + \int_0^1  dx [\rho_A(x) \ln \rho_A(x)
+\rho_B(x) \ln \rho_B(x) +\rho_C(x) \ln \rho_C(x)]  \nonumber \\
&& - \beta \int_0^1 dx \int_0^1 dz{  \sin(c({1 \over 2} - z )) \over 2 \sin({c  \over 2 } )}
[  \rho_A(x) \rho_B(x+z)  +\rho_B(x) \rho_C(x+z) +\rho_C(x) \rho_A(x+z)] \nonumber \\
&& - 3 \beta^2 \int_0^1 dx \int_0^1 dz{  \cos(c({1 \over 2} - z )) \over 4
 c \sin({c  \over 2 } )} [  r_C(1-3 r_C)\rho_A(x) \rho_B(x+z)
\nonumber  \\  
&& \qquad \qquad \qquad +r_A(1-3 r_A)\rho_B(x) \rho_C(x+z) +r_B(1-3 r_B)\rho_C(x) \rho_A(x+z)] \nonumber
\end{eqnarray}
where
$$c= \beta \sqrt{\Delta - 9 r_A r_B r_C} $$
was compatible both with   (\ref{ld2}) at order $\beta^2$ for arbitrary
density fluctuations and with the exact expressions of the correlation
functions (\ref{rara4},\ref{rarb4}), i.e. for arbitrary $\beta$ but small
deviations of the density from the flat profiles.  What this expression
becomes for general $\beta $ and arbitrary deviations  of the density is an open question.

\vskip1cm

The macroscopic fluctuation theory developed by Bertini et al. \cite{BDGJL1,BDGJL2} (see also \cite{FW}) 
relates the steady state large deviation functional (\ref{F-def}) to the dynamical large deviation functional 
(\ref{pro}-\ref{eq: I ABC}). 
One can try to apply this general procedure to the ABC model.
We fix $0< \beta < \beta_c$ such that the steady state profiles are flat (equal to $r_A,r_B$) and such that the hydrodynamic equations (\ref{eq: hydro}) relax to $r_A,r_B$ for any initial density profiles (this property excludes situations with metastable profiles).
Then the  density large deviation functional (\ref{F-def})  is given by 
\begin{eqnarray}
{\cal F}[\rho_A(x), \rho_B(x), \rho_C(x)] = \inf_{\hat \rho, \hj} \left\{ {\cal I}_{[-\infty,0]} (\hat \rho, \hj) \right\} \, ,
\label{eq: quasi pot ABC}
\end{eqnarray}
where the infimum of the functional ${\cal I}_{[-\infty,0]}$ (\ref{eq: I ABC}) is taken over the trajectories  $\big( \hat \rho (x,t), \hj (x,t) \big) = \{ \hat \rho_A (x,t), \hat \rho_B (x,t), \hj_A (x,t), \hj_B (x,t) \}$ starting at time $-\infty$ from the steady state configurations 
$(r_A,r_B)$ and ending at time $t=0$ at the configuration $(\rho_A(x), \rho_B(x))$ (with $\rho_C(x) = 1 - \rho_A(x) - \rho_B(x)$).
To recover the large deviation functional in the steady state, there is no other constraint on the currents $\hj$ besides the conservation law $\partial_t \hat \rho = - \partial_x \hj$.
Finding the optimal trajectory is in general an open problem which has been solved only in rare instances \cite{BDGJL2,BGL}.
For the sake of completeness, the equations satisfied by the optimal trajectories are presented in Appendix II.

A perturbative analysis of the variational problem (\ref{eq: quasi pot ABC}) leads to an expansion of ${\cal F}$ in the vicinity of the steady state. For example (\ref{ld3}) can be obtained as the second order expansion of ${\cal F}$ and the two-point correlations can then be recovered by inverting the quadratic form in  (\ref{ld3}).
Expanding ${\cal F}$ at the second order is essentially equivalent to the mode decomposition implemented in this section (\ref{perturb}). To see this, we first note that (\ref{eq: quasi pot ABC}) can be rewritten as a variational principle on the time interval $[-\infty,\infty]$ with the constraint that the trajectory $\hat \rho$ is equal to $(\rho_A(x), \rho_B(x))$ at time 0
\begin{eqnarray}
{\cal F}[\rho_A(x), \rho_B(x), \rho_C(x)] = \inf_{\hat \rho, \hj} \left\{ {\cal I}_{[-\infty, \infty]} (\hat \rho, \hj) \right\} \, .
\label{eq: quasi pot ABC 2}
\end{eqnarray}
This identity comes from the fact that the contribution for positive times is null as one can choose $\hat \rho$ which relaxes to equilibrium 
according to (\ref{eq: hydro}). For small deviations, ${\cal I}_{[-\infty, \infty]} (\hat \rho, \hj)$ can then be expanded with respect to the Fourier modes (\ref{perturb}).
The different modes decouple and we recover a quadratic form with coefficients related to the correlations (\ref{eq: quadratic}). The constraint on the density at time 0 is then achieved by optimizing this quadratic form over the temporal modes $\omega$.
This leads to a second order expansion of ${\cal F}$ (in terms of the spatial Fourier modes) from which the steady state two-point correlation functions (\ref{rara2}-\ref{rarb2}) can be obtained.


\section{Diffusive systems on a ring}

The second example we analyze is the phase transition \cite{BDGJL2006,BD2005,BD2007}  which is expected to occur in the large deviation function of the current in some diffusive systems such as  the weakly asymmetric exclusion process (WASEP).

The WASEP on a ring is defined as follows: one considers a fixed number
$N=L r$ of particles on a ring of $L$ sites, with at most one particle per site.  Each particle hops to its neighboring site on its right with rate $q^{-1}$ and to its neighboring site on its left with rate $q$, provided that the target site is empty. As for the ABC model, we consider here the diffusive regime where $q$ scales with the system size $L$ as in (\ref{q-scaling})
$$ q = \exp\left[-{\beta \over L}\right] \, .$$
For general diffusive systems  with one species of particles on a ring, one can write  
an expression very similar to (\ref{pro},\ref{H-def}) for the probability of observing a density profile $\rho(x,\tau)$ and a (rescaled)  current $j(x,\tau)$ over
a (rescaled) time interval $T$ (the microscopic time $t= T L^2$)
\begin{equation}
 {\rm Pro}(\rho,j) \sim \exp \left( -L  \ \I_{[0,T]} (\rho,j) \right)
\label{pro1}
\end{equation}
where 
\begin{equation}
\I_{[0,T]} (\rho,j) =  \int_0^1 dx \int_0^T d \tau   \ {\cal H} (\rho,j) \, ,
\label{eq: func}
\end{equation}
with
 \begin{equation}
{\cal H}(\rho,j)= {[j(x,\tau) + D(\rho(x,\tau)) {d \rho(x,\tau) \over dx} - \beta \sigma(\rho(x,\tau))]^2 \over 2 \sigma(\rho(x,\tau)} \  .
\label{H-def1}
\end{equation}
and the current $j$ and the density $\rho$ are related, as usual,  by the conservation law
\begin{equation}
{d \rho \over d \tau} = - {d j \over dx} \, .
\end{equation}
We refer to \cite{BD2007, BDGJL2006}  for further details on the derivation of (\ref{eq: func}).
The functions $D(\rho)$ and $\sigma(\rho)$ in (\ref{H-def1}) are two
functions characteristic of the diffusive system. 
For the WASEP, as defined above, these two functions are $D(\rho)=1$ and $\sigma(\rho)=2
\rho (1- \rho)$. 
In general their ratio is related to the compressibility at equilibrium by the Einstein relation
(see  \cite{kubo} equation (4.6.1) or \cite{HS}).

If one considers the flux $Q_t(i)$  of particles between sites $i$ and $i+1$ during a long microscopic time $t$,
the  large deviation function $G(j_0)$ of the current describes the distribution of $Q_t(i)$,
in the long time limit
\begin{equation}
{\rm Pro} \left( {Q_t(i) \over t} = {j_0 \over L} \right) \sim \exp\left[ -{t \over L} G(j_0) \right]
\label{G-def}
\end{equation}
(the $L$ dependence which appears in (\ref{G-def}) can be understood
because the system is diffusive \cite{BD2007, BDGJL2006}).
For systems with conservative dynamics such as the WASEP, which are irreducible Markov processes with a finite number of states, the large deviation function $G(j_0)$ depends neither on the initial condition at time $0$ nor on the final configuration at time $t$ nor on the section $i$ chosen.

In what follows we  consider (instead of  the flux $Q_t(i)$ through a section) the total flux $Q_t$
\begin{equation}
Q_t= \sum_{i=1}^L Q_t(i)
\label{Qt-def}
\end{equation}
which has the same large deviation function
\begin{equation}
{\rm Pro} \left( {Q_t \over t} = {j_0 } \right) \sim \exp\left[ - {t \over L} G(j_0) \right] \ .
\label{G-def1}
\end{equation}

According to the macroscopic fluctuation theory, one can calculate
this large deviation function $G(j_0)$ by looking at the time evolution of the density and of the current which maximizes (\ref{pro1}) with the constraint that
\begin{equation}
{1 \over T} \int_0^T d \tau \int_0^1 dx  \ j(x,\tau) = j_0
\label{j0}
 \end{equation}
so that
\begin{equation}
G(j_0) = \lim_{T \to \infty}  {1 \over T} \left\{\min_{\rho(x,\tau),j(x,\tau)}
\ \I_{[0,T]} (\rho,j)
\right\}
\label{variational}
\end{equation}

When 
 the optimal  density  and current profiles in (\ref{variational}) are constant in time and in space
\cite{BDGJL2006}, 
the expression of $G(j_0)$ follows immediately 
\begin{equation}
G(j_0) = {[j_0 - \beta \sigma(r)]^2 \over 2 \sigma(r)}
\label{G-expr}
\end{equation}
where $r=N/L$ is the density of particles along the ring.

By analyzing the neighborhood of the flat profile solution,
it was shown in \cite{BD2005} that this profile    is locally unstable
when
\begin{equation}
8 \pi^2 D^2(r) \sigma(r) +[\beta^2 \sigma^2(r) - j_0^2] \sigma''(r) <0
\label{local-stability}
\end{equation}
and the optimal profiles becomes space or space and time dependent.
(Note that, as for the $ABC$ model, (\ref{local-stability}) is obtained by a local stability analysis.
Thus the optimal profiles might already be space or space and time dependent  even when  (\ref{local-stability}) is not satisfied with the occurrence of first order transitions).

\medskip

We are now going to see that, although there is no trace  of the second order phase transition  (\ref{local-stability}) in  the expression (\ref{G-expr}), the correlation functions become singular along the stability  line (\ref{local-stability}).
There are several ways of  defining the correlation functions, conditioned on the current: if the flux is conditioned to be $Q_t = j_0 t  $ over a time interval $-t/2< t' < t/2$, one expects  the correlation $ \langle \rho(x,t') \rho(y,t') \rangle_c $  to depend on whether $t'$ is close or far from the boundaries of the time interval $(-t/2,t/2)$. We are now going to calculate two of these correlation functions
\begin{equation}
\label{cor-1}
\langle \rho(x) \rho(y) \rangle_{\rm intermediate} = \lim_{ 
t \to \infty}\langle \rho(x,0) \rho(y,0) | Q_{t}=j_0 t \rangle_c
\end{equation}
\begin{equation}
\label{cor-2}
\langle \rho(x) \rho(y) \rangle_{\rm final} =  \lim_{t\to \infty} \langle \rho(x,{t / 2}) \rho(y,{t / 2}) | Q_{t}=j_0 t\rangle_c
\end{equation}

During the macroscopic time interval $[-T/2,T/2]$,
one considers a small perturbation of the flat profile $r$ of the form 
\begin{eqnarray}
\label{eq: approx traj}
\rho(x,\tau) = r + k [f_1(\tau) \cos(k x) + f_2(\tau) \sin(k x)] \\
j(x,\tau) = I_0(\tau)  + f_2'(\tau) \cos(k x) - f_1'(\tau) \sin(k x) \nonumber
\end{eqnarray}
where $f_1(\tau)$ and $f_2(\tau)$ are  a priori two arbitrary functions and $k=2 \pi n$ 
with $n \ge 1$, integer).
The parameter $I_0(\tau)$ takes into account the constraint on the current in the time interval $[- T/2, T/2]$
\begin{eqnarray*}
\left \lbrace
\begin{array}{l}
I_0(\tau) = j_0, \qquad  {\rm if} \ -T/2 < \tau < T/2  \,, \\
I_0(\tau) = \beta \sigma (r), \qquad  {\rm otherwise} \, .
\end{array}
\right.
\end{eqnarray*}
One can rewrite (\ref{H-def1}) to quadratic order
\begin{equation}
{\cal H}(\rho,j)= A(\tau) (f_1^2 + f_2^2) +B(\tau) ({f_1'}^2+{f_2'}^2) +C(\tau) (f_1 f_2' - f_2 f_1') + E(\tau) (f_1 f_1'+f_2 f_2') 
\label{H-def2}
\end{equation}
where
\begin{eqnarray}
&& A(\tau) = {D^2 k^4 \over 4 \sigma} + {\beta^2 k^2 \sigma'' \over 8} +
 {k^2 \sigma'^2 I_0(\tau)^2 \over 4 \sigma^3}
 -{k^2 \sigma'' I_0(\tau)^2 \over 8 \sigma^2}
\nonumber \\ 
&& B(\tau) = {1 \over 4 \sigma}, \quad C(\tau) = -{k \sigma' I_0(\tau)  \over 2 \sigma^2},
\quad E(\tau) = {D k^2     \over 2 \sigma} 
\label{ABCE} 
\end{eqnarray}
with the functions $D,\sigma,\sigma',\sigma''$  evaluated at $\rho=r =
N/L$.

When $T$ goes to infinity, the expectations $\langle f_1^2(0)\rangle, \langle f_1(0) f_1(0)
\rangle,\langle f_2^2(0)\rangle $ can then be computed  in the following two situations (see Appendix I):
\begin{itemize}
\item in the  case (\ref{cor-1}), one has  $I_0(\tau)=j_0$ for all times
$\tau$. One gets from (\ref{eq: A1})
\begin{equation}
\label{res1}
\langle f_1(0)^2 \rangle = \langle f_2(0)^2 \rangle = {\sigma   \over   L  \sqrt{D^2 k^4  + {\beta^2 k^2 \sigma \sigma'' \over 2} - {k^2 \sigma'' j_0^2 \over 2 \sigma}}} 
\end{equation}
and by summing the contributions of all the modes,  this gives
\begin{equation}
\label{cor-1-bis}
\langle \rho(x) \rho(y) \rangle_{\rm intermediate} =  { 2 \pi  \over    L}
\sum_{n \ge 1} {n  \cos(2 \pi n (y-x)) \ \sigma 
\over  \sqrt{ 4 \pi^2 n^2 D^2   + {\beta^2  \sigma \sigma''
\over 2} - { \sigma'' j_0^2 \over 2 \sigma}}} \, .
\end{equation}
\item in the case (\ref{cor-2}), one has $I_0(\tau) =j_0$ for $\tau \le 0$ and $I_0(\tau) = \beta \sigma(r)$ for $\tau >0$
\begin{equation}
\label{res2}
\langle f_1(0)^2 \rangle = \langle f_2(0)^2 \rangle = {2 \sigma   \over    L
\left[ D
k^2 +  \sqrt{D^2 k^4  + {\beta^2 k^2 \sigma \sigma'' \over 2} - {k^2
\sigma'' j_0^2 \over 2 \sigma}} \  \right]} 
\end{equation}
and the sum over all the modes leads to 
\begin{equation}
\label{cor-2-bis}
\langle \rho(x) \rho(y) \rangle_{\rm final} =  
{ 4  \pi  \over    L}
\sum_{n \ge 1} {n  \cos(2 \pi n (y-x)) \ \sigma
\over  2 \pi n D + \sqrt{ 4 \pi^2 n^2 D^2   + {\beta^2  \sigma \sigma''
\over 2} - { \sigma'' j_0^2 \over 2 \sigma}}} \, .
\end{equation}
\end{itemize}
We see in expressions (\ref{cor-1-bis},\ref{cor-2-bis}) that the
correlation functions become singular along the transition line
(\ref{local-stability}),  where the $n=1$ mode  becomes unstable.
Note that only the intermediate correlation function (\ref{cor-1-bis}) diverges at the transition.

\bigskip

Let $\bar j_0 = \beta \sigma(r)$ be the mean current, then one can rewrite the correlation function 
(\ref{cor-1-bis}) as
\begin{eqnarray}
\label{cor-1-bis bis}
&& \langle \rho(x) \rho(y) \rangle_{\rm intermediate} = \\
&& \qquad   {\sigma  \over  2  L  D}
\sum_{n \ge 1}  2 \cos(2 \pi n (y-x))  
\left( 1 + \left[ {1 \over  \sqrt{ 1 - { {\sigma''   \over 8 \sigma \pi^2 n^2 D^2  } (j_0^2 - (\bar j_0)^2} )  }} - 1 \right] \right) \, .
\nonumber 
\end{eqnarray}
When there is no constraint on the current then $j_0 = \bar j_0$ and (\ref{cor-1-bis bis}) reduces (see (\ref{eq: cos identite})) to 
\begin{eqnarray*}
\langle \rho(x) \rho(y) \rangle_{\rm intermediate} =  
{ \sigma  \over   2 L D} \sum_{n \ge 1} 2 \cos(2 \pi n (y-x))  
= { \sigma  \over   2 L D} \big( \delta (x-y) -1 \big)  \, ,
\end{eqnarray*}
which are the correlations for the (micro-canonical) invariant measure of the diffusive part of the dynamics.
This was already emphasized in \cite{BDGJL-cor, BDGJL5}.
On the other hand, if the system is conditioned to an atypical current deviation, the term in the bracket  in (\ref{cor-1-bis bis}) does not vanish.
It is interesting to note that all the modes have the same sign which depends on $j_0^2 - \bar j_0^2$ and $\sigma''$.
One can check that depending on the sign of $ j_0^2 - \bar j_0^2$ the correlations for $x-y$ small are either positive or negative: 
 the particles tend to cluster or to spread according to the constraint on the current.
We interpret this clustering phenomenon as a precursor of the macroscopic clustering which occurs after the transition.

Instead of the perturbation  (\ref{eq: approx traj}), one could also have used a space/time decomposition over the Fourier modes as in (\ref{perturb}). 
However this would have led to extra difficulties to treat the fluctuations at the final time. In the latter case,
the functional depends on the current constraint which is not uniform in time so that the contributions of the modes 
at different frequencies would be coupled (unlike the case of intermediate fluctuations).
In the particular case of the WASEP, a microscopic approach based on a truncation procedure as in section
\ref{sec: truncation} would also have led to the correlations (\ref{cor-1-bis}).

In \cite{ADLW}, it was shown recently that a very similar phase
transition occurs in the SSEP (the symmetric exclusion process) when one
considers the large deviation function of the activity $K_t$ (which is the
total number of changes of configurations during time $t$). 
A calculation almost identical
to the one presented in this section can be done to calculate the
correlation functions, conditioned on the value of $K_t$. 
The  nature of the singularities of the correlation functions
   are then  very similar to those found in (\ref{cor-1-bis}, \ref{cor-2-bis}).

\section{Conclusion}

In the present work, we have calculated, using two different approaches,  the long range correlation functions in the
$ABC$ model and shown that they become singular at the second order phase transition
(\ref{rara1},\ref{rarb1},\ref{rara2},\ref{rarb2},\ref{rara4},\ref{rarb4}).
We have also calculated, for general diffusive systems, the two point correlation functions, conditioned on the current
(\ref{cor-1-bis}, \ref{cor-2-bis}) and seen that they become singular at a second order phase transition. Similar calculations can be done, when conditioned on other quantities such as the activity \cite{ADLW}.

It would be interesting to try to extend our results to other phases than the flat phase or to other situations than the ring geometry, like open systems.
The large scale Gaussian fluctuations of the density, which here  are at the origin of the long range correlations, allow also to calculate all the cumulants of the current \cite{ADLW} for diffusive systems. It would be interesting to see whether one could establish  more direct or  more general relations between the long range correlations and the distribution of the current, and to know how these relations are modified in the case of open systems.

Lastly we have obtained an improved expression (\ref{ld3}) of the large
deviation functional of the density for the $ABC$ model. This expression is still an approximation. It would be interesting to go further, for example by calculating higher correlations using the truncation procedure (\ref{sca2},\ref{sca3}).

\vskip1cm

\noindent
{\it Acknowledgments.}
TB and BD acknowledge the support of
the French Ministry of Education through the ANR BLAN07-2184264 grant.
FvW acknowledges the support of the French Ministry of Education through the ANR-05-JCJC-44482 grant.

\section*{Appendix I}

In this appendix,  we establish the expressions (\ref{res1},\ref{res2}).
Let us consider two Gaussian functions  $h_1(\tau),h_2(\tau)$  of the time $\tau$ distributed according to  the following distribution
\begin{eqnarray}
&& {\rm Pro}(\{h_1(\tau),h_2(\tau)\}) \sim  \exp\left[ -L \int_{-\infty}^\infty d \tau [A(\tau) 
  (h_1^2+ h_2^2) + B(\tau) (h_1'^2+ h_2'^2) \right. \nonumber \\
&& \qquad \qquad  \qquad   \left. + C(\tau) (h_1 h_2' - h_1' h_2) + E(\tau) (h_1h_1'+h_2 h_2')
\right]
\label{gaussian} 
\end{eqnarray}
Here we are interested in situations where the functions $A(\tau),B(\tau),C(\tau), E(\tau)$  take some constant values $A_-,B_-,C_-,E_-$ for $\tau <0$  and other constant values (possibly the same)  $A_+,B_+,C_+,E_+$ for $\tau >0$.

If we fix $h_1(0)$ and $h_2(0)$, and one  integrates (\ref{gaussian})  over
$h_1(\tau)$ and $h_2(\tau)$  for all times $\tau \ne 0$ one gets
$$
{\rm Pro}(\{h_1(0),h_2(0)\}) \sim
\exp\left[ -L  \left(\alpha_+ B_+  + \alpha_-  B_- + {E_--E_+ \over 2} \right) (h_1(0)^2 + h_2(0)^2) \right]
$$
where $$\alpha_+=  {\sqrt{4 A_+ B_+ - C_+^2} \over 2 B_+}
 \ \ \ \ \ ; \ \ \ \ \alpha_-=  {\sqrt{4 A_- B_- - C_-^2} \over 2 B_-} $$

In fact it can be shown that the functions which maximize  (\ref{gaussian})  for  $\tau>0$ are, at fixed $h_1(0) ,h_2(0)$   
\begin{eqnarray}
\label{eq: trajectory}
h_1(\tau) = e^{-\alpha_+ \tau} [ h_1(0) \cos(\beta_+ \tau) + h_2(0) \sin(\beta_+ \tau) ]\\
h_2(\tau) = e^{-\alpha_+ \tau} [ h_2(0) \cos(\beta_+ \tau) - h_1(0) \sin(\beta_+ \tau) ] \nonumber
\end{eqnarray} 
where $$\beta_+= {C_+  \over 2 B_+} \ .$$
Similar expressions with $\alpha_+ $ and $\beta_+$ replaced by $ - \alpha_-$ and $ \beta_-$ give these optimal functions for $\tau <0$.
This leads to
\begin{eqnarray}
\label{eq: A1}
\langle h_1(0)^2 \rangle = \langle h_2(0)^2 \rangle = {1  \over [2 (\alpha_+  B_+ + \alpha_- B_-) + E_--E_+] L } 
\end{eqnarray} 
and $\langle h_1(0) h_2(0) \rangle =0$.


\section*{Appendix II}

In this Appendix, we study the variational problem (\ref{eq: quasi pot ABC}). We set 
\begin{eqnarray*}
\cond(\hat \rho)=   
\left( 
\begin{array}{cc}
   2   \rho_A (1-\rho_A) & - 2 \rho_A \rho_B\\
      - 2 \rho_A \rho_B & 2 \rho_B (1-\rho_B) \\
   \end{array} \right)
   \qquad
   F(\hat \rho) = 
 \left(  \begin{array}{c} 
     \rho_A (\rho_A + 2 \rho_B -1) \\
     \rho_B (1- 2\rho_A - \rho_B) \\
   \end{array} \right)
\end{eqnarray*}
The trajectories are denoted by $\hat \rho= (\rho_A, \rho_B)$ and the currents $\hj = (j_A, j_B)$.  
It will be convenient to rewrite the currents in terms of the new variables $H=(H_A, H_B)$
\begin{eqnarray}
\label{eq: current}
\hj(x,t) = 
-  \partial_x \hat \rho(x,t) - \beta F (\hat \rho(x,t)) + \cond(\hat \rho(x,t)) \partial_x H(x,t) \, , 
\end{eqnarray}
with $H(0,t)=H(1,t)=0$.
In particular
\begin{eqnarray*}
\partial_t \hat \rho(x,t) = 
\partial^2_x \hat \rho(x,t)  + \beta \partial_x \Big( F(\hat \rho(x,t) \Big) -   \partial_x \Big( \cond (\hat \rho (x,t)) \partial_x H(x,t)) \Big) \, .
\end{eqnarray*}
Thus the functional (\ref{eq: I ABC}) reads 
\begin{eqnarray}
\label{eq: I ABC 2}
{\cal I}_{[-\infty,0]} \big( \hat \rho, \hj \big) = \frac{1}{2}
\int_{-\infty}^0 dt  \int_0^1 dx  \;  
\big( H_A ' , H_B ' \big)  
 \ \cond(\rho(x,t)) \ 
\left( \begin{array}{c} 
H_A '   \\
H_B ' 
\end{array}  \right) \, ,
\end{eqnarray}
where $H_A', H_B'$ stand for the spatial derivatives.
Optimizing the functional implies that the optimal trajectories satisfy 
\begin{eqnarray}
\label{eq: opt traj}
\partial_t \rho_A  &=& 
\Delta \rho_A + \beta \partial_x \Big( \rho_A (\rho_A + 2 \rho_B -1) \Big) 
-  2 \partial_x \Big( \rho_A (1-\rho_A)  H_A' -  \rho_A \rho_B H_B' \Big)  \nonumber \\
\partial_t \rho_B  &=& 
\Delta \rho_B   + \beta \partial_x \Big( \rho_B (1- 2\rho_A - \rho_B)  \Big) 
-  2 \partial_x \Big( \rho_B (1-\rho_B)  H_B' -  \rho_A \rho_B H_A' \Big) \\
\partial_t H_A  &=&  -  \Delta H_A
+ \beta  (2 \rho_A + 2 \rho_B -1) H_A' - 2 \beta \rho_B H_B' - \Big( (1-2\rho_A) (H_A')^2 - 2 \rho_B H_A' H_B' \Big)
 \nonumber \\
\partial_t H_B  &=&  -  \Delta H_B
+ 2 \beta \rho_A  H_A' - \beta   (2 \rho_B + 2 \rho_A -1) H_B' - \Big( (1-2\rho_B) (H_B')^2 - 2 \rho_A H_A' H_B' \Big)
 \nonumber
 \end{eqnarray}
with the constraint on the densities at time 0, $\big( \rho_A (x,0), \rho_B(x,0) \big)= \big(\rho_A (x), \rho_B(x) \big)$ and 
$(r_A, r_B)$ at time $-\infty$.

\bigskip

Solving (\ref{eq: opt traj}) amounts to knowing the functional ${\cal F}$ (see \cite{BDGJL2} equation (2.16)) 
\begin{eqnarray*}
 H_A  = {\partial {\cal F} \over \partial \rho_A}, \qquad  H_B  = {\partial {\cal F} \over \partial \rho_B} \, .
\end{eqnarray*}
In the case $r_A = r_B = 1/3$, the optimal drifts can be guessed from the explicit expression  (\ref{ld1}) of the functional ${\cal F}$
\begin{eqnarray*}
&& H_A(x) = \log\left( {\rho_A(x) \over 1 - \rho_A(x) - \rho_B(x)} \right)
+ \beta \left( \int_0^x du \, \rho_B(u) -\rho_A(u) + \right. \\
&& \qquad \qquad \left.
\int_x^1 du \, \big( 1- \rho_B(u) -\rho_A(u) \big)
+ \int_0^1 du \, (u-x) \big( 3 \rho_B(u) - 1 \big) \right)
\end{eqnarray*}
and
\begin{eqnarray*}
H_A' (x) = \frac{(\rho_B(x)-1) \rho_A '(x)- \rho_A(x) \rho_B'(x)}{\rho_A(x) (\rho_A(x)+\rho_B(x)-1)} +
\beta \Big( 3 \rho_B(x)-1 \Big) \, .
\end{eqnarray*}
A similar expression holds for $H_B$ leading to
\begin{eqnarray*}
\partial_t \rho_A  &=& - \Delta \rho_A(x,t) - \beta \partial_x \Big( \rho_A (\rho_A + 2 \rho_B -1) \Big) \, , \\
\partial_t \rho_B  &=& - \Delta \rho_B(x,t) - \beta \partial_x \Big( \rho_B (1- 2 \rho_A - \rho_B) \Big) \, .
\end{eqnarray*}
Thus the optimal trajectories are the time reversed of the hydrodynamic equations (\ref{eq: hydro}). This could have been guessed from the microscopic reversibility derived in \cite{CDE}. 
Under the assumption that the hydrodynamic equations (\ref{eq: hydro}) relax to the flat densities $r_A = r_B =1/3$, the expression  (\ref{ld1}) of ${\cal F}$ can be recovered
from the optimal trajectories by integrating (\ref{eq: I ABC 2}) over the time.

\end{document}